\documentclass[journal]{IEEEtran}

\usepackage[utf8]{inputenc}
\usepackage{graphicx}
\usepackage{amsmath}
\usepackage{subfigure}
\usepackage{color}
\usepackage{soul}

\begin{document}

\title{Green Adaptation of Real-Time Web Services for Industrial CPS within a Cloud Environment}

\author{M. Teresa Higuera-Toledano,
        José L. Risco-Martín,
        Patricia Arroba,
        and José L. Ayala
\thanks{M. Teresa Higuera-Toledano, José L. Risco-Martín, and José L. Ayala are with the Department of Computer Architecture and Automation, Complutense University of Madrid, C/Prof. José García Santesmases 9, 28040 Madrid, Spain, email: {mthiguer,jlrisco,jlayalar}@ucm.es}
\thanks{Patricia Arroba is with the Department of Electronic Engineering, Technical University of Madrid, Avda. Complutense 30, 28040, Madrid, Spain, email: parroba@die.upm.es}
\thanks{Copyright (c) 2009 IEEE. Personal use of this material is permitted. However, permission to use this material for any other purposes must be obtained from the IEEE by sending a request to pubs-permissions@ieee.org.}}

\markboth{Journal of \LaTeX\ Class Files,~Vol.~11, No.~4, December~2016}%
{Higuera-Toledano \MakeLowercase{\textit{et al.}}: Bare Demo of IEEEtran.cls for Journals}

\maketitle

\begin{abstract}
Managing energy efficiency under timing constraints is an interesting
and big challenge. This work proposes an accurate power model in data
centers for time-constrained servers in Cloud computing. This model,
as opposed to previous approaches, does not only consider the workload
assigned to the processing element, but also incorporates the need of
considering the static power consumption and, even more interestingly,
its dependency with temperature. The proposed model has been used in a
multi-objective optimization environment in which the \emph{Dynamic
  Voltage and Frequency Scaling} (DVFS) and workload assignment have
been efficiently optimized.
\end{abstract}


\section{Introduction}

\IEEEPARstart{B}{oth} \emph{Cyber Physical Systems} (CPSs) and \emph{Cyber Physical Society} \cite{Lee:2010:CF:1837274.1837462} combine computing and networking power with physical components, enabling innovation in a
wide range of domains related to future-generation sensor networks
(e.g., robotics, avionics, transportation, manufacturing processes,
energy, smart homes and vehicles, medical implants, healthcare,
etc). The design and implementation of CPS involve the consideration
of multiple aspects like energy and tight real-time
constraints. Because of that, real-time scheduling for CPS brings new
research issues in the scope of real-time systems \cite{journals/pieee/BanerjeeVMG12}.

Managing energy efficiency under timing constraints is a big
challenge. Most modern micro-controllers already provide support for
various energy saving modes (e.g., Intel Xeon and AMD Opteron). A
common way of reducing dynamic power is to use the technique called
\emph{Dynamic Voltage and Frequency Scaling} (DVFS), which changes the
processor voltage and the clock frequency simultaneously, reducing the
energy consumption. Decreasing the processor voltage and frequency
will slow down the performance of the processor. If the execution
performance is not a hard constraint, then, decreasing both processor
voltage and frequency allows to reduce the dynamic power consumption
of the processor.

Nowadays, new embedded devices are collaborating in distributed environments. In this new scenario, tasks and resources are widely distributed and then, real-time applications become more complex and more relevant.  A cloud datacenter usually contains
a large group of servers connected through the Internet, and a scheduler has to make an efficiently use of the resources of the cloud to execute jobs.
Since many applications require \emph{Quality of Service} (QoS), power
consumption in data centers must be minimized, satisfying the
\emph{Service Level Agreement} (SLA) constraints.
Consequently, novel approaches that base their optimizations on
accurate power models must be devised, performing an optimized setting
of the parameters of the server (frequency, voltage, workload
allocation, etc) while accomplishing with time requeriments and a wide
range of real-time constraints.

DVFS-based solutions for distributed real-time environments identify
two main dimensions of the problem: \emph{(i)} task-to-\emph{Central-Processing-Unit} (CPU) allocation
and \emph{(ii)} run-time voltage scaling on individual CPUs. In CPS,
physical factors (e.g., the network topology of CPS may
  dynamically change due to physical environments) are not entirely
predictable and may lead to problems such as missed task deadlines,
that can impact dramatically on economic loss for individuals or
for the industry. Moreover, a critical task deadline missed could
trigger a disaster (e.g., humans life loss, natural disasters, or huge
economic loss).

In this paper, we propose a method for solving such CPS problems by
introducing new adaptive real-time scheduling algorithms in
distributed computing infrastructures that also consider energy
efficiency. This scheme requires to know \emph{a priori} the
processing and timing constraints of the set of tasks, and must be
supported by reservation-based real-time operating systems.

The remainder of this paper is organized as follows: after a brief
summary of the previous works in this field
(Section~\ref{sec:related}), a real-time scheduling algorithm for CPS
is sketched (Section~\ref{sec:Job-model}). Following, the devised
power model is presented (Section~\ref{sec:model}), and the
optimization of the algorithm developed is profusely described
(Section~\ref{sec:milp}). Experimental results can be found in
Section~\ref{sec:results}. Finally, some conclusions are drawn
(Section~\ref{sec:conclusions}).

\section{Related work}\label{sec:related}

The energy-efficient scheduling problem in real-time systems consists
in minimizing the energy consumption while ensuring that all the
real-time tasks meet their deadlines. The work presented in
\cite{Rowe:2008:RSS:1475690.1475895} is based on the observation that
a significant percentage of time spent in idle mode is due to the
accumulation of small gaps between tasks. Whether the gap between the
activation of two periodic tasks is less than transition-time from
idle to deep-sleep, the processor is not able to transition to the
deep-sleep state even though there is no useful work to be done, and
continues in the idle energy state all the time.

There are extensive research works on energy-aware real-time
scheduling by using DVFS (e.g., \cite{DBLP:conf/rtcsa/ChenK07}).
Different works using this technique within a real-time context, considered an offline scheduling algorithm and a set of a periodic jobs on an ideal processor. Each job is characterized by its release time, deadline, and execution CPU cycles, and all jobs have the same power consumption function.

Several papers have also proposed DVFS-based solutions for real-time
multi-processor systems. As the complexity of CPS increases,
\emph{Chip Multicore Processors} (CMP) and parallel tasks scheduled in
a real-time way are needed. The fact that the processing cores share a
common voltage level makes the CMP energy-efficiency problem different
from multi-processor platforms. The work presented in
\cite{DBLP:conf/rtcsa/KandhaluKLR11} provides a simple, elegant and
effective solution on energy-efficient real-time scheduling on
CMP. This solution addresses fixed-priority scheduling of periodic
real-time tasks having a deadline equal to their period. Note that
this problem is NP-hard.

The load balancing in CMP is particularly important because the main
contributor to the overall energy consumption in the system is the core
with the maximum load. This fact is given by the global
voltage/frequency constraint. Considering a CMP system with a workload
perfectly balanced across the processors, the \emph{Earliest Deadline
  First} (EDF) scheduling minimizes the total energy consumption. This
is not the case of \emph{Rate Monotonic Scheduling} (RMS) where
load-balancing does not always result in lowering energy consumption
\cite{DBLP:conf/rtcsa/KandhaluKLR11}.

In mixed-criticality systems, varying degrees of assurance must be
provided to functionalities of varying importances. As shown in
\cite{Huang:2014:EED:2656045.2656057} there is a conflict between
safety and energy minimization because critical tasks must meet their
deadlines even whether exceeding their expected \emph{Worts Case
  Execution Time} (WCET).  This work integrates continuous DVFS with
the EDF with \emph{Virtual Deadlines} (EDF-VD) scheduling for mixed-criticality systems
\cite{conf/ecrts/BaruahBDLMSS12} and shows that speeding up the system
to handle overrun is beneficial for minimizing the expected energy
consumption of the system.

Generally, large-scale distributed applications require real-time
responses to meet soft deadlines. Hence, the middleware coordinates
resource allocation in order to provide services accomplishing with
SLA requirements. In \cite{Wang2010}, we can find a scheduling
algorithm based on DVFS for clusters, which develops a green SLA-based
mechanism to reduce energy consumption by increasing the scheduling
makespans. In \cite{Beloglazov:2012:ERA:2148243.2148369}, we can find
an energy-aware resource allocation for Cloud computing with
negotiated QoS. However, similarly to the solution presented in
\cite{Wang2010}, this method sacrifices system performance.

The work presented in \cite{DBLP:journals/fgcs/WuCC14} proposes a
priority-based scheduler, which satisfies the minimum resource
requirement of a job by selecting a \emph{Virtual Machine} (VM) according to both the SLA level
and the $W_{i}$ parameter that is described as $W_{i} = P_{i} \times
R_{i}$, where $P_{i}$ is the unit power cost of $VM_{i}$, and $R_{i}$
defines the resources used by the $VM_{i}$.

The location of nodes in CPS affects the effective release time and
deadline of real-time tasks, which may be different depending on the
node location and the migration delay time among the network
nodes. Because of that, traditional real-time scheduling algorithms
have to be modified to include the location node and the spatial
factors. The work presented in \cite{DBLP:journals/fgcs/ParkKF14}
proposes a CPS scheduling algorithm, where the servicing node (i.e., the CPU) needs to move to serviced (i.e., the executed Job) node for
real-time services.

The power modeling technique proposed in \cite{Mobius2014} is most
relevant for us. A correlation between the total system's power
consumption and the component utilization is observed, defining a
four-dimensional linear weighted power model for the total power
consumed (i.e.,
$P=c_{0}+c_{1}P_{CPU}+c_{2}P_{cache}+c_{3}P_{DRAM}+c_{4}P_{disk}$). Our
work follows a similar approach but also incorporates the contribution
of the static power consumption, its dependency with temperature, and
the effect of applying DVFS techniques.

Static power consumption has a high impact on energy, due to the temperature-dependent leakage
currents.  In this manner, novel optimizations may be devised by
quantitatively understanding the power-thermal trade-offs of a system,
thus developing analytical models.

Finally, Rafique et al.~\cite{Rafique:2011:PMH} makes a description of
the complexity of the power management and allocation
challenge. Authors demonstrate that achieving an optimal allocation
depends on many factors as the server’s maximum and minimum frequencies,
the job’s arrival rate, and consequently, the relationship between power
and frequency.  They conduct a set of experiments that provides
significant savings in terms of energy in both homogeneous and
heterogeneous clusters. However, our work presented in this paper
outperforms these savings by exploiting a multi-objective optimization
strategy to help to minimize the servers’ power for time-constrained
Cloud applications.

\section{The industrial services execution model}\label{sec:Job-model}

CPS comprise a large number of sensors and actuators, and
  computing units that exchange different types of data, some of these
  interactions have real-time constraints. Real-time system
  abstraction and hybrid system modeling and control are among the CPS
  research challenges.  The hybrid system model of CPS requires the
  design and integration of both the physical and computational (i.e.,
  cyber) elements. While physical elements behave in continuous
  real-time, computational elements change according to discrete
  logic. This fact requires to merge continuous-time based systems
  with event-triggered logical systems, and also we must address the
  dimensional scale (i.e., from on-chip level to the cloud).
  Moreover, the interaction with physical world introduces uncertainty
  in CPS because of randomness in the environment, errors in physical
  devices, and security attacks.

Control and scheduling co-design is a well-known area in the embedded
real-time systems' community. However, since CPS are typically
networked control systems, the tradeoff between the effects of the
network must be included in the real-time schedulability, that results
in a non-periodic control approach. In this work, we study how to
guarantee the overall system stability with minimum computational
resource and power usage. System properties and requirements
(e.g., the control laws, real-time and power constraints) must be
captured and supported by data abstractions encapsulated in
components.

\subsection{Task characterization}

Typically CPS's are composed of hard real-time tasks and feedback
control tasks. Whereas real-time tasks present time constraints (i.e.,
deadlines) that must always be satisfied, feedback control tasks are
characterized by their \emph{Quality of Control} (QoC), which needs to be optimized. A typical
approach to the above scheduling problem is to translate the QoC
requirements into time constraints and then, to apply traditional
real-time scheduling techniques \cite{Davis2011}. Real-time systems are structured as a set of schedulable tasks,
  where parameters used for the scheduling (e.g., execution time,
  deadline, or period) are a priori known and clearly defined.
  However, this solution is very conservative and consequently it is
  not efficient for CPS.

An alternative solution is the given in \cite{Masrur2013d}, that deals
with this problem using a \emph{multi-layered} scheme based on
mixed-critical real-time systems: \emph{(i)} for real-time tasks it
uses triggering patterns (i.e., uses arrival curves), which allow a
more general characterization regarding the classical real-time task
models (i.e., \emph{periodic} or \emph{sporadic}), and \emph{(ii)} for
control tasks, it is based on three QoC-oriented
metrics. Mixed-critical real-time systems literature focuses on tasks
with different criticality levels and certification
issues\footnote{When there are tasks with different
    safety requirements into the same real-time platform, it is called
    mixed-criticality system.}, providing heterogeneous
  timing guarantees for tasks of varying criticality levels.

As an example, in the \emph{Unmanned Aerial Vehicles} (UAVs),
functionalities can be categorized as safety-critical tasks (e.g.,
like flight control and trajectory computation) or mission-critical
tasks (e.g., object tracking for surveillance purposes). Note that the
system is still safe although mission-critical functionalities can be
lost.  This makes the design parameters for safety-critical tasks
(e.g., WCET) much more pessimistic than those for mission-critical
tasks.  However, in CPS, tasks are not characterized by criticality
levels, but by their criticality types.

 There has been considerable research on schedule
   synthesis for control applications. However, these works are
   particularly centered on control/scheduling co-design for optimized
   QoC, and only deal with control tasks.  On the other hand, CPS
   focus on mixed task sets comprising of feedback control tasks and
   hard real-time tasks, which requires a joint schedule synthesis.  

\subsection{The task model}

In CPS, tasks may be classified according to their
  criticality types (e.g., deadline-critical real-time tasks and
  QoC-critical feedback control tasks).
While the system must satisfy always the deadlines of real-time tasks,
particularly for those that are critical, only the QoC parameters for
control tasks need to be optimized.  In order to do that, we require
stochastic hybrid systems to identify the interaction between
continuous dynamical physical models and discrete state machines, and
the CPS architecture must follow the new paradigm \emph{``globally
  virtual, locally physical''}.

\begin{figure}[!h]
  \includegraphics[width=0.47\textwidth]{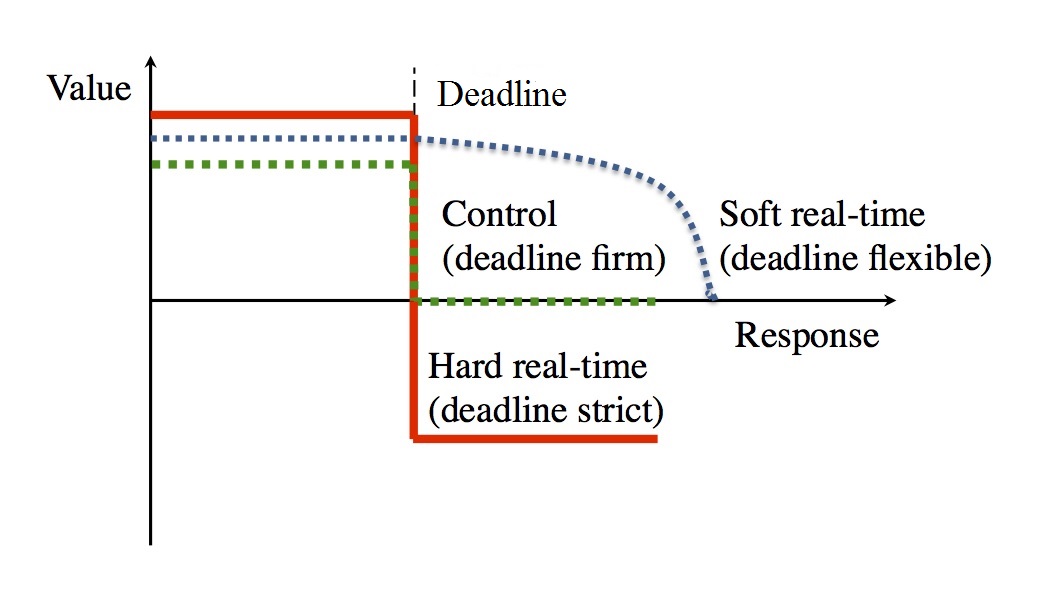}
  \caption{An overrun in response time (i.e., a deadline miss) has a different value function depending on its possible consequences}  
  \label{fig:Deadline}
\end{figure}

We consider a set of independent tasks, (i.e., $\Sigma$) which are
executed remotely in a set of physical servers $m$. We define our
real-time problem as a pair $P = (\Sigma,S)$ where $S$ is a scheduling
solution and $\Sigma = {\tau_{1},...,\tau_{n}}$ is a set of ${n}$
tasks with different timing characteristics (i.e., strict, flexible,
and firm) as shows Figure \ref{fig:Deadline}.

\noindent
Each task $\tau_{i}$ is a possibly infinite sequence of jobs (i.e.,
demands for processing time), each one with an associated
deadline. Jobs of the same task must be executed sequentially and in First-In-First-Out (FIFO) order. If the timing characteristics of the task $\tau_{i}$ are
soft or firm, the jobs may be not identical.

CPS requires jointly scheduling hard real-time, soft
  real-time or best-effort, and control-feedback tasks. Due to the
  stringent stability requirements, we classify control tasks as firm
  deadline.  While a hard deadline cannot be missed, soft deadlines
  may be occasionally missed and it does not harm the system
  safety. Similarly, firm deadlines can be missed but there is an
  upper limit on the number of misses within a given time interval.
  However, as we aim to optimize the QoC, we must minimize the number
  of deadline misses to avoid QoC degradation.  The characterization
  of each type of task is fundamentally different as follows.

\subsubsection{Hard real-time tasks} 
A real-time system is considered \emph{hard} if an overrun in a task
response time leads to potential loss of life and/or big financial
damage. The system is considered to be safety critical or high
integrity, and is often mission critical. We consider a real-time task
as a tuple $\tau_{i} = (R_{i}, C_{i}, T_{i}, D_{i})$ where:

\begin{description}
\item[$R_{i}$] is the first release time of the task (i.e. the \emph{phase}
of the task), 
\item[$C_{i}$] is the WCET,
\item[$T_{i}$] is the activation period (i.e., minimum inter-release time), and
\item[$D_{i}$] is the relative deadline ($r_{i} \leq D_{i} \leq T_{i})$. The absolute deadline is the relative deadline plus the current arrival time.
\end{description}

\noindent We compute the CPU utilization factor of $\tau_{i}$ as $U_{i}=\frac{C_{i}}{T_{i}}$.

\subsubsection{Soft real-time tasks} 
For \emph{soft} real-time tasks, deadline overruns are tolerable but
not desired (i.e., there are not catastrophic consequences of missing
one or more deadlines). There is a cost function associated with these
systems, which is often related to QoS. Hence, we
consider a stochastic task model based on the one presented in
\cite{Abeni2012}. Then, we represent each soft-real-time task using a
tuple $\tau_{i} = (r_{i}, s_{i}, a_{i}, d_{i})$ where:

\begin{description}
\item[$r_{i}$] is the release time of the task,
\item[$s_{i}$] is the service time, which follows an exponential
  distribution of average $\mu^{-1}$ (i.e., $\mu$ is the number of
  serviced jobs of $ \tau_{i}$ per unit time),
\item[$a_{i}$] is the arrival time; tasks arrive according to a
  renewal process with exponential distribution of average
  $\lambda^{-1}$, and
\item[$d_{i}$] is the absolute deadline; the relative deadline is
  $D_{i} = d_{i} - a_{i}$, $D_{i}$ distributed on $[0, D]$.
\end{description}

\noindent We compute the response time of $\tau_{i}$ as $\rho_{i} =
c_{i}-a_{i}$, where $c_{i}$ is the completion time (i.e., $c_{i} =
a_{i}+s_{i}$). The average CPU utilization factor is given by
$\Upsilon_{i}=\frac{\mu_{i}}{\lambda_{i}}$.

\subsubsection{Feedback control tasks} 
For a \emph{firm} real-time task the computation is obsolete whether
the job is not finished on time. In this case, the cost function may
be interpreted as loss of value associated to QoC. This is the case of
the feedback control task in CPS. For this kind of task we can
consider $D_{i} \geq T_{i})$ to guarantee that the controlled physical
tasks are still stable in the worst case scenario. However, this sacrifices the system
performance and also may result unstable under physical perturbations.

In most cases, feedback control systems become unstable with too many
missed control cycles. Therefore, a critical question is how to
determine $T_{i}$ to ensure both schedulability and robustness of the
physical system. Considering a simple proportional-gain
feedback controller, which is fixed for each control task, in order to
determine $T_{i}$, we can find the minimum $T_{i} \in (T_{1}, T_{2},
\ldots, T_{n})$ under the following constraints:
\begin{eqnarray}
 0 \leq & \sum_{i}\frac{ C_{i}}{T_{i}} & \leq p \\
C_{i} \leq & T_{i} & \leq D_{i} 
\end{eqnarray} 

\noindent where $p < 1$ is a priori known. However, some controller
parameters may need to be adjusted when the task period is changed.
Alternatively, we can use a multiple-versions approach or a predictive
model with a quadratic optimization computed iteratively for each
job. However, very often, probabilistic guarantees are sufficient
(e.g., $t$ out of $k$ deadlines have to be met).

Permitting skips in periodic tasks increases the system flexibility
\cite{Chantem2009}. The maximum number of skipped jobs for each task
can be controlled by a specific parameter $S_{i}$ associated with the
task, which gives the minimum distance between two consecutive jobs
skips (e.g., if $(S_{i} = 3)$ the task can skip one job every
three). This parameter can be considered as a QoC metric (i.e., the
higher S, the better QoC).

When $S_{i} = \infty$ no skips are allowed, meaning that $\tau_{i}$ is
a real-time hard periodic task. We then consider a control task as a
tuple $\tau_{i} = (R_{i}, C_{i}, T_{i}, D_{i}, S_{i})$ where $T_{i} =
D_{i}$.

\subsection{The parallel  scheduling}

For each of the above described tasks $\tau_{i} \in \Sigma$, we
consider a set of independent subtasks $\tau_{i} =
{\tau_{i,1},...,\tau_{i,q}}$, where $\tau_{i,j}$ denotes the subtasks
$j$ of task $\tau_{i}$. Therefore, $e_{i} \geq 0$, is the energy
consumption rate of the task $\tau_{i}$ per time unit:
\begin{eqnarray}
e_{i} & = & \bigcup_{j=1}^{|\tau_{i}|} e_{i,j}
\end{eqnarray}

\noindent The scheduling allocates each $\tau_{i,j}$ subtask in a set
of $m$ physical servers, taken into account the critical timing
characteristics of each task $\tau_{i}$ and the minimal energy
consumption of the task set $\Sigma$.

The performance criteria generally used in systems when the model task
does not have explicit deadlines, is to minimize the task delay (i.e.,
the response time of all tasks). However, when there are explicit
deadlines, we must ensure that critical tasks fulfill their deadline
and minimize the fraction of non-critical tasks that do not meet their
timing requirements.

We can consider lateness constraints of the form $\alpha (x) \leq
\beta$, where $\alpha(x)$ is the fraction of jobs that miss
their deadline by more than $x$ time units. Here, missing a deadline
by $x$ time units is considered as a failure.

\begin{itemize}

\item For firm deadlines, we require that $\alpha(0) \leq \beta$
  (i.e., the fraction of tasks missing their deadliness were limited
  to $\beta$).  Note that this has a different meaning for the $S$
  parameter, which is the minimal distance between the consecutive
  misses of the task $\tau_{i}$. Hence, we consider a $\tau_{i}$
  missing whether one or more subtasks $\tau_{i,j}$ of a job miss the
  deadline.

\item For hard real-time tasks, we establish $\alpha(0) \leq 0$ (i.e.,
  we do not tolerate any deadline missed), while for each control task
  $\tau_{i}$, $\alpha(0) \leq \frac{S_{i} - 1}{S_{i}}$.

\item For soft real-time tasks, we generalize, $\alpha(x_{i}) \leq
  \beta_{i}$, for a set of time values ${ x_{1}, ..., x_{p}}$ and
  constraint specifications ${ \beta_{1}, ..., \beta _{p}}$, where $1
  \leq i \leq p$, which allows to take into account the stochastic
  nature of task arrivals and service time of soft real-time tasks.

\end{itemize}

\section{Power and energy model}\label{sec:model}

Traditionally in electronic systems, dynamic consumption has been the
major contributor to the power budget. In contrast, when scaling
technology below $100nm$, static consumption reaches the $30-59\%$ of
the total power, thus becoming much more
significant~\cite{narendra2010leakage}. Moreover, the exponential
impact of temperature on leakage currents intensifies this effect. Thus,
modeling leakage will allow the exploitation of the trade-offs between
leakage and temperature at the server level when taking decisions on
resource configuration and selection.

Therefore, the impact of static consumption must be considered, taking
into account its correlation with temperature. This section presents
our leakage-aware static power model. We validate this model using
real data gathered from real machines of our case study (e.g., Intel
Xeon and AMD Opteron).

\subsection{Leakage power}
Equation~\eqref{eq:eq14} shows the impact of leakage on the currents
in a MOS device.  Rabaey demonstrates in his work that, when $V_{DS} >
100 mV$, the second exponential may be considered
negligible~\cite{rabaey2009low}.  Consequently, the previous equation
may be revised as in \eqref{eq:eq15}, also regrouping technological
parameters together obtaining the formula presented in
equation~\eqref{eq:eq18}.

\begin{eqnarray}
I_{leak} & = & I_{s} \cdot e^{ \frac{V_{GS} - V_{TH}}{nkT/q}} \cdot (1 - e^{\frac{Vds}{kT/q}}) \label{eq:eq14} \\
I_{leak} & = & I_{s} \cdot e^{\frac{V_{GS} - V_{TH}}{nkT/q}} \label{eq:eq15} \\
I_{leak} & = & B \cdot T^2 \cdot e^{\frac{V_{GS} - V_{TH}}{nkT/q}} \label{eq:eq18}  
\end{eqnarray}

The leakage power consumption for the physical machine $m \in \{1,
\ldots, M\}$ presented in Equation~\eqref{eq:eq20} can be inferred
from the expression in~\eqref{eq:eq19}. Then, the expansion of the
mathematical expression in its Taylor $3rd$ order series provides
Equation~\eqref{eq:eq21}, where $B_{m}$, $C_{m}$ and $D_{m}$ represent
the technological constants of the server.

\begin{eqnarray}
P_{leak, m} & = & I_{leak, m} \cdot V_{DD, m} \label{eq:eq19} \\
P_{leak, m} & = & B_{m} \cdot T_{m}^2 \cdot e^{\frac{V_{GS} - V_{TH}}{nkT/q}} \label{eq:eq20} \\
P_{leak, m} & = & B_{m} \cdot T_{m}^2 \cdot V_{DD, m} \nonumber \\
            & + & C_{m} \cdot T_{h} \cdot V_{DD, m}^2 + D_{m} \cdot V_{DD, m}^3 \label{eq:eq21} 
\end{eqnarray}


\subsection{Dealing with DVFS}

The main contributors to energy consumption in nowadays servers are
CPU and memory devices. Despite DVFS is easily found in CPUs, there
are still few memories with these capabilities.  However, memory
consumption in some cases (memory-intensive applications) is very
significant compared to the CPU consumption and, because of this, it
was considered important enough to be studied independently.

Equation~\ref{eq:eq119} provides the consumption of a physical server
that has $k \in \{1 \ldots K\}$ DVFS modes, while memory remains at a
constant voltage. This expression takes into account the impact of
temperature on the static power contribution. We define $E_{m}$ as the
contribution of other server resources operating at constant values of
frequency and voltage.

\begin{eqnarray}
P_{leak, mk} & = & B_{m} \cdot T_{CPU, m}^2 \cdot V_{DD, mk} \nonumber \\
             & + & C_{m} \cdot T_{CPU, m} \cdot V_{DD, mk}^2 + D_{m} \cdot V_{DD, mk}^3 \nonumber \\
             & + & E_{m} + G_{m} \cdot T_{MEM, m}^2 + H_{m} \cdot T_{MEM, m} \label{eq:eq119}
\end{eqnarray}

In order to measure temperature-dependent leakage we must understand
also the dynamic contribution of the server’s power consumption. To
maintain constant conditions, we use
\emph{lookbusy \footnote{http://www.devin.com/lookbusy/}}, which is a
synthetic application that stresses the CPU during specifics periods
of time. \emph{Lookbusy} is able to stress, not only the cores but
also the hardware threads of the CPU at a precise utilization, having
no impact on memory or disk devices. Synthetic workloads help us to
maintain the utilization rate constant (in terms of instructions per
cycle), thus revealing the leakage contribution due to temperature
variations. The formulation of the dynamic power consumption is shown
in Equation~\ref{eq:dynpwr}.

\begin{eqnarray}
P_{CPU, dyn, imk} & = & A_{m} \cdot V_{DD, mk}^2 \cdot f_{mk} \cdot u_{CPU, imk} \label{eq:dynpwr}  
\end{eqnarray}

\noindent where $A_{m}$ defines the technological constant of the
physical machine $m$ and $f_{mk}$ and $V_{DD, mk}$ are respectively
the frequency and the supply voltage at the $k$ DVFS mode of the
CPU. $u_{CPU, imk}$ represents the CPU utilization and it is
correlated with the number of CPU cycles.


\subsection{Energy model}

So far, the power model is derived as in \eqref{eq:totpwr}.
\begin{eqnarray}
P_{tot, mk} & = & A_{m} \cdot V_{DD, mk}^2 \cdot f_{mk} \cdot \sum_{i}{u_{CPU, imk}} \nonumber \\
           & + & B_{m} \cdot T_{CPU, m}^2 \cdot V_{DD, mk} \nonumber \\
           & + & C_{m} \cdot T_{CPU, m} \cdot V_{DD, mk}^2 \nonumber \\
           & + & D_{m} \cdot V_{DD, mk}^3 + E_{m} \label{eq:totpwr}
\end{eqnarray}

The corresponding energy model can be easily obtained taking into
account that $E=P \times t$, being $P$ the power model in
\eqref{eq:totpwr} and $t$, the execution time. Thus, the total energy
consumed per host is described as the summation of the following
equations:

\begin{eqnarray}
E_{CPU, dyn, mk} & = & A_{m} \cdot V_{DD, m}^2 \cdot CPI \nonumber \\
                 &   & \cdot \sum_{i}{u_{CPU, imk} \cdot n_{CPU, imk}} \label{eq:Ecpu} \\
E_{leak, mk} & = & [ \nonumber \\
             &   & B_{m} \cdot T_{CPU, m}^2 \cdot V_{DD, m} \nonumber \\
             & + & C_{m} \cdot T_{CPU, m} \cdot V_{DD, m}^2 + D_{m} \cdot V_{DD, m}^3 \nonumber \\
             & + & E_{m} + G_{m} \cdot T_{MEM, m}^2 + H_{m} \cdot T_{MEM, m} \nonumber \\
             &   & ] \cdot \frac{CPI}{f_{mk}} \cdot \sum_{i}{n_{CPU, imk}} \label{eq:leak} 
\end{eqnarray}

\noindent where
\begin{itemize}
\item $CPI$ is the number of cycles per instruction
\item $n_{CPU, imk}$ is the number of CPU instructions of each task
  $i$ assigned to be executed in a specific server $m$ and DVFS mode
  $k$.
\end{itemize}

The summation of both the instructions to execute and the resources
used by the workload hosted on the server are needed in order to get
the execution time of all tasks executed in parallel considering the
resources offered by each server, as seen in \eqref{eq:leak}.
\begin{eqnarray}
E_{tot} & = & \sum_{mk} {(E_{CPU, dyn, mk} + E_{leak, mk})} \label{eq:totener}       
\end{eqnarray}

\section{Multi-Objective Optimization Algorithm}\label{sec:milp}
In this work, we aim for a workload allocation in a cloud that allows
to optimize energy consumption.  In addition, the benefits offered by
virtualization are exploited, allowing to allocate the tasks in a more
versatile way. The proposed system is defined as a cluster of machines
of a cloud facility.

The proposed solution considers server heterogeneity, so the
technological parameters will vary from one architecture to another,
resulting in a different energy consumption.  Since the resultant
power model is non-linear and there exists a large set of constraints,
the problem is tackled as a multi-objective optimization:

\begin{eqnarray}
\mathrm{Minimize} & & \nonumber \\
\mathbf{y} = \mathbf{f}(\mathbf{x}) & = & \left[\lambda, (1+\lambda) \cdot E_{tot}(\mathbf{x})\right] \nonumber \\
\mathrm{Subject~to} & & \nonumber \\
\mathbf{x} & = & (x_1,x_2,\ldots , x_n) \in \mathbf{X} \label{eq:multiobj}
\end{eqnarray}

\noindent where $\mathbf{x}$ is the vector of $n$ decision variables,
$\mathbf{f}$ is the vector of 2 objectives function, $\lambda$ is the
number of constraints not satisfied, $E_{tot}$ is the total energy,
and $\mathbf{X}$ is the feasible region in the decision space. Using
$\lambda$ as shown in Equation~\ref{eq:multiobj}, unfeasible solutions
are also allowed, but only when no other alternatives are found. In
this particular case, $E_{tot}$ is measured using \eqref{eq:totener},
whereas $\lambda$ is computed as a penalization over the control and
soft tasks that are delivered after the deadline (see Figure
\ref{fig:Deadline}).

Using this formulation, we are able to obtain optimal energy savings,
realistic with the current technology. To provide an efficient
assignment in data centers it is necessary to consider both the energy
consumption and the resource needs of each task of the workload.

A task $\tau_{i}$ can be split in different subtasks $\tau_{i,j}$ in
order to achieve energy savings. Therefore, a task $\tau_{i}$ can be
executed using a specific amount of resources of one or more servers
defined by $u_{CPU, imk}$. The utilization percentage of the resources
assigned to a task determines its execution time (i.e., $C_{i}$ or
$s_{i}$). In summary, the proposed multi-objective formulation, once
solved, decides the following aspects:

\begin{itemize}
\item Operating server set, indicating which hosts are active
  according to the operating conditions of each physical machine.
\item Best assignment for the various tasks of the workload,
  distributing each CPU instruction and memory requirements according
  to the minimum energy consumption of the applications in the
  computing infrastructure. For control tasks, $S=2$ must be
  fulfilled. However, a penalty is added to $\lambda$ when one
  control task is aborted, even when $S$ is being satisfied.
\item Percentage of resources used by every task in each host where it
  is allocated, achieving best energy consumption.
\end{itemize}


\subsection{The solver}
Evolutionary algorithms have been used to run the proposed
multi-objective formulation. In this work, we use the Non-dominated
Sorting Genetic Algorithm II (NSGA-II) \cite{Deb2002}, which has
become a standard approach to solve this kind of problems
\cite{Sayyad2013}. The chromosome encoding is shown in
Figure~\ref{fig:chromosome}.

\begin{figure}[!h]
  \centering
\begin{small}
  \begin{tabular}{|c|c|c|c|c|c|}
  \hline
  $DVFS_{1}$ & $\cdots$ & $DVFS_{M}$ & $n_{CPU, 11}$ & $\cdots$  &   $n_{CPU, NM}$ \\
  \hline
  \end{tabular}
\end{small}
  \caption{Chromosome encoding}
  \label{fig:chromosome}
\end{figure}

\noindent In this case, each gene represents a decision
variable. Because many decision variables are integer, the chromosome
uses integer encoding. Decision variables like $n_{CPU, imk}$ are
scaled to the integer interval $0 \leq n_{CPU, imk} \leq 100$, and
transformed to its real value (i.e., multiplying the percentage by the
total number of instructions in the multi-objective function for
evaluation).

NSGA-II is always executed with an initial random population of $100$
chromosomes. After that, the algorithm evolves the population applying
(1) the NSGA-II standard tournament operator, (2) a single point
crossover operator with probability of $0.9$ as recommended in
\cite{Deb2002}, (3) a integer flip mutation operator (with probability
of $1/\mathrm{number~of~decision~variables}$ as also recommended in
\cite{Deb2002}, and (4) the multi-objective evaluation. Steps (1) to
(4) are applied for a variable number of iterations or generations,
which depend on the time that the parameter $\lambda$ becomes $0$
(usually $25000$ iterations have been enough).


\section{Results}\label{sec:results}

Tests have been conducted gathering real data from a Fujitsu RX300 S6
server based on an Intel Xeon E5620 processor and a SunFire V20z Dual
Core AMD Opteron 270, both operating at the set of frequencies
$f_{mi}$ given in Table~\ref{tab:Frequencies}. Total power consumption
and CPU temperature have been collected via the \emph{Intelligent Platform Management Interface} (IPMI) during the execution
of \emph{lookbusy} at different utilization levels ranging from 0\% to
100\%, where a 65\% of these levels were used to fit the
  energy model and the remaining 35\% for validation. We used MATLAB
to fit our data, obtaining the constants and validation errors shown
in Table~\ref{tab:fittingconstants}.

\begin{table}[!h]
\centering
\caption {Intel Xeon E5620 and SunFire V20z Dual Core AMD Opteron 270
  frequencies}
\begin{tabular}{lllllll}
\hline
Platform & $f_{m1}$ & $f_{m2}$ & $f_{m3}$ & $f_{m4}$ & $f_{m5}$ & $f_{m6}$ \\
\hline
Intel (GHz) & 1.73 & 1.86 & 2.13 & 2.26 & 2.39 & 2.40 \\
\hline
AMD (GHz) & 1.0 & 1.8 & 2.0 & & & \\
\hline
\end{tabular}
\label{tab:Frequencies}
\end{table}

\begin{table*}[!t]
  \centering
  \caption {Constants obtained for Power curve fitting}
\begin{tiny}
  \begin{tabular}{lcccccccccccccc}
  \hline
  Server & $A$ & $B_{1}$ & $B_{2}$ & $C_{1}$ & $C_{2}$ & $D$ & $E$ & $F$ & $G_{1}$ & $G_{2}$ & $H_{1}$ & $H_{2}$ & Error & Temp. range\\
  \hline
  Intel & 	14.3505 & 0.1110 & -	   & -0.0011 & -      & 0.3347 & -40700 & 64.9494 & 275.702 & 	-	  & -0.4644 & - & 11.28\% & 293-309K\\
  AMD  &	11.2390 & 1.9857 & -6.1703 & -0.0002 & 0.0132 & 426.51 & -5.3506 & 25.1461 & -444.480 & 464.076 & 0.6977 & -0.7636 & 9.12\% & 293-312K\\
  \hline
  \end{tabular}
\end{tiny}
  \label{tab:fittingconstants}
\end{table*}

The efficiency of the power supplies affects the calculation of these
constants for different temperatures. In consequence, negative
constants appear due to the fact that only CPU and memory have been
characterized in this work because of their dominant contribution. In
order to adapt the problem to more generic Cloud computing
environments, our model constants can be calculated for data obtained
during the execution of the workload in virtual machines. In that
experimental approach, both the power model and the multi-objective
optimization formulations would still be valid.

Once the model proposed in section~\ref{sec:model} for both Intel Xeon
and AMD Opteron servers have been validated, we have proceeded with
the analysis of results. The considered performance parameters are the
temperature of both CPU and memory, as well as the frequency and
voltage of the DVFS modes available to the CPU in each physical
machine. These variables modify independently the dynamic and static
consumption of servers in each architecture, so different behaviors
for Intel and AMD have been found. Table \ref{tab:tasksProfile} shows
the set of tasks used for the optimization.

\begin{table}[!htb]
\centering
\caption {Profile of tasks allocated}
\begin{small}
\begin{tabular}{lcrccc}
\hline
Task Id & Type & \# Ins & Period & Deadline & \# Jobs \\
\hline
0 & REAL & 7740796 & 114.20 & 0.021 & 131 \\
1 & CTRL & 5594832 & 114.21 & 0.015 & 115 \\
2 & REAL & 4138643 & 137.12 & 0.011 & 112 \\
3 & CTRL & 98156923 & 124.66 & 0.267 & 95 \\
4 & REAL & 739437676 & 124.76 & 2.01 & 118 \\
5 & SOFT & 2591877 & 124.86 & 0.007 & 103 \\
6 & SOFT & 3093531 & 124.85 & 0.008 & 112 \\
7 & SOFT & 5447445 & 105.76 & 0.015 & 115 \\
8 & SOFT & 5722568 & 152.21 & 0.016 & 99 \\
\hline
\end{tabular}
\end{small}
\label{tab:tasksProfile}
\end{table}

These tasks have been adapted from the TUDelft workloads
archive\footnote{http://gwa.ewi.tudelft.nl/datasets/gwa-t-1-das2}. The task set consists of a number of deadline-critical tasks
  $\tau_{hrt}=\{\tau_{0}, \tau_{2}, \tau_{4}\}$, a number of
  QoC-critical control tasks $\tau_{c}=\{\tau_{1}, \tau_{3}\}$, and a
  number soft real-time tasks $\tau_{srt}=\{\tau_{5}, \tau_{6},
  \tau_{7}, \tau_{8}\}$. We assume that all tasks are independent from
  each other.  However, due to the interference from other tasks, each
  task $\tau_{i}$ experiences a response time or delay $R_{i}$.  
Periods and deadlines are given in seconds. Each real-time tasks $\tau_{hrt}$ is bounded to one single host. Only control $\tau_{c}$ and soft tasks $\tau_{srt}$ are
allowed to loss their deadline, increasing the $\lambda$ parameter in
the multi-objective function. Control tasks are configured with $S=2$.

NSGA-II has been executed with the minimum frequency in all the
CPUs (labeled in the results as DVFS-MIN), the maximum frequency
(labeled as DVFS-MAX) and a range of 5 possible DVFS modes (from 1 to
5). This algorithm has been compared with a more traditional approach, the EDF-VD algorithm. The overall goal is to design a priority assignment
  technique with the following objectives:
\begin{itemize}
\item 
All the real-time tasks $\tau_{hrt}$ meet their deadlines $D_{hrt}$ in
the WCET
\item The overall QoC of all the control tasks $\tau_{c}$ and QoS of
  all the soft real-time tasks $\tau_{srt}$ is maximized.
\item The overall energy is minimized.
\end{itemize}

\begin{figure}
  \centering
  \includegraphics[width=0.75\columnwidth]{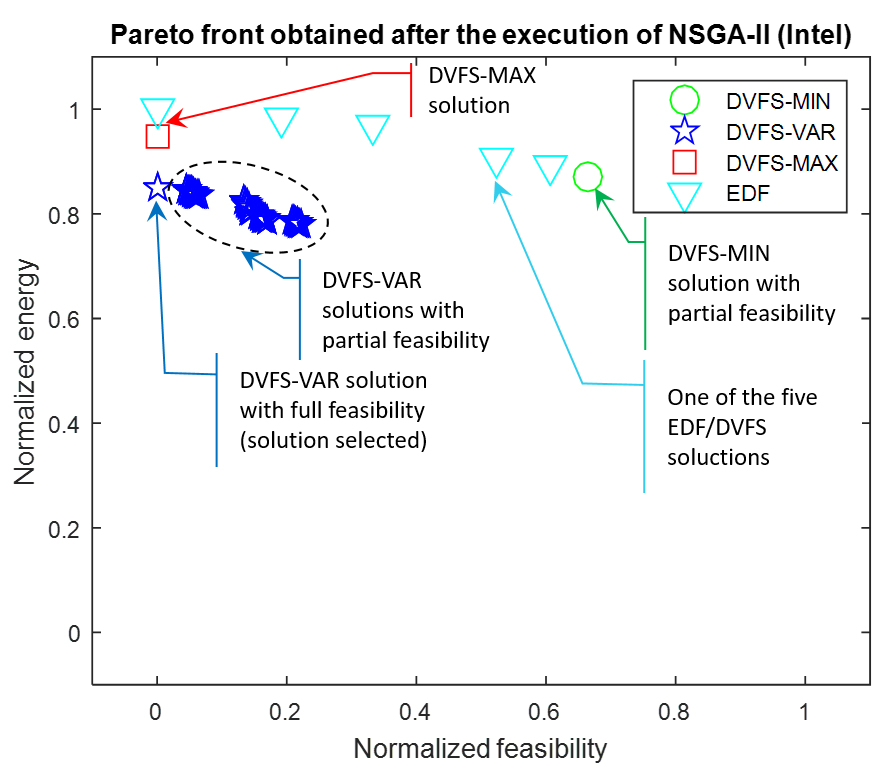}
  \caption{Pareto front obtained with NSGA-II after optimizing the
    allocation of tasks over the Intel architecture.}
  \label{fig:IntelFront}
\end{figure}

Figure \ref{fig:IntelFront} depicts the three obtained
  Pareto fronts for the Intel architecture. Both objectives have been
  normalized to the worst value in all the Intel and AMD
  optimizations (1 unit of energy = 95.6 KJ). As can be seen, the DVFS-MAX Intel framework is able
  to allocate all the tasks in Table \ref{tab:tasksProfile} without
  penalizations (labeled as full feasibility). Using DVFS-MIN, the
  algorithm was not able to allocate all the required tasks, having to
  break some soft timing constraints (labeled as partial
  feasibility). As can be seen, there is at least one DVFS-VAR
  configuration able to execute all the tasks without penalization and
  with less energy than DVFS-MAX and close to DVFS-MIN. Table
  \ref{tab:AmdDvfs-IntelDvfs} shows the DVFS modes selected by the
  DVFS-VAR solution with full feasibility.

\begin{figure}
  \centering
  \includegraphics[width=0.75\columnwidth]{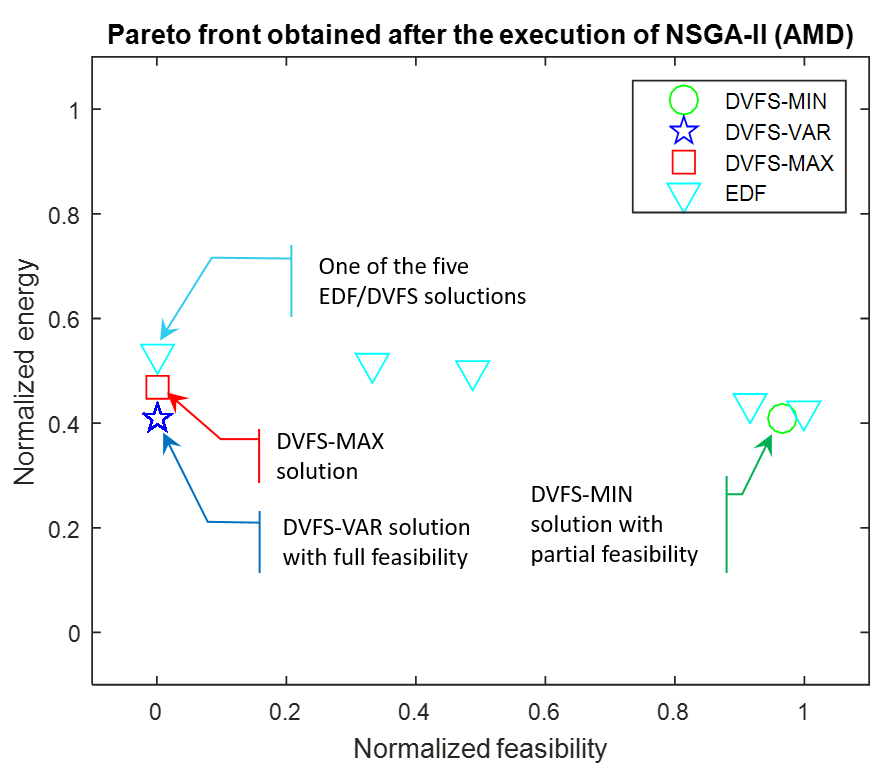}
  \caption{Pareto front obtained with NSGA-II after optimizing the
    allocation of tasks over the AMD architecture.}
  \label{fig:AmdFront}
\end{figure}

Similarly, Figure \ref{fig:AmdFront} shows the three obtained
non-dominated fronts for the AMD architecture. As with the Intel
scenario, the algorithm was not able to execute all the REAL, CTRL and
SOFT tasks without penalization using the minimum DVFS mode
(DVFS-MIN), although all the REAL tasks were properly
executed. However, we found a completely feasible solution in DVFS-VAR
(feasibility=0), consuming less energy than DVFS-MAX. Table
\ref{tab:AmdDvfs-IntelDvfs} shows the DVFS models selected by the
multi-objective algorithm in the DVFS-VAR AMD optimization.

EDF was able to schedule all the tasks in both cases, but using the maximum DVFS mode and thus consuming more energy than the proposed algorithm.

\begin{table}[!h]
\centering
\caption {DVFS modes obtained by NSGA-II Pareto front in the DVFS-VAR
  optimization}
\begin{tabular}{lllllll}
\hline
Platform & CPU 1 & CPU 2 & CPU 3 & CPU 4 & CPU 5 & CPU 6 \\
\hline
Intel & 2 & 2 & 2 & 2 & 2 & 5 \\
\hline
AMD & 5 & 1 & 5 & & & \\
\hline
\end{tabular}
\label{tab:AmdDvfs-IntelDvfs}
\end{table}

As a result, the best DVFS configuration that can execute all the
demanded services given in Table \ref{tab:tasksProfile} has been found
without penalizations, obtaining a high diversity in terms of energy
consumption.

\section{Conclusions}\label{sec:conclusions}

CPS and Mobile Cloud Computing have collided with the
lack of accurate power models for the energy-efficient provisioning of
their devised infrastructures, and the real-time management of the
computing facilities.  In this paper, we have presented a
  reservation-based scheme aiming to jointly schedule
  deadline-critical, QoS non-critical, and QoC tasks. The work
proposed in this paper has made substantial contributions in the area
of power modeling of high-performance servers for Cloud computing
services under timing constraints, which is an interesting and big
challenge.

We have proposed an accurate power model in data centers for time
constrained servers in Cloud computing, which does not only consider
the workload assigned to the processing element, but also incorporates
the need of considering the static power consumption and its
dependency with temperature.

The proposed model has been used in a multi-objective optimization
environment in which the DVFS and workload assignment have been
efficiently optimized in a realistic scenario composed of Fujitsu
RX300 S6 servers based on an Intel Xeon E5620 and SunFire V20z Dual
Core AMD Opteron 270. Results show that the proposed
  multi-objective optimization framework is able to find the best DVFS
  configuration that can execute all the given demanded services
  without penalizations. In addition, the set of non-dominated
  solutions found presents a high diversity in terms of energy
  consumption.

The obtained results open a motivating research line that could enable
intensely sought Green computing paradigm with hard timing
constraints.  Future work envisages to extend the
  scheduling model to integrate the concept of criticality levels.  

\section*{Acknowledgment}
This work is supported by the Spanish Ministry of Economy and Competitivity under research grants TIN2013-40968-P and TIN2014-54806-R.

\bibliographystyle{IEEEtran}
\bibliography{Bibliography}

\begin{thebibliography}{10}
\providecommand{\url}[1]{#1}
\csname url@samestyle\endcsname
\providecommand{\newblock}{\relax}
\providecommand{\bibinfo}[2]{#2}
\providecommand{\BIBentrySTDinterwordspacing}{\spaceskip=0pt\relax}
\providecommand{\BIBentryALTinterwordstretchfactor}{4}
\providecommand{\BIBentryALTinterwordspacing}{\spaceskip=\fontdimen2\font plus
\BIBentryALTinterwordstretchfactor\fontdimen3\font minus
  \fontdimen4\font\relax}
\providecommand{\BIBforeignlanguage}[2]{{%
\expandafter\ifx\csname l@#1\endcsname\relax
\typeout{** WARNING: IEEEtran.bst: No hyphenation pattern has been}%
\typeout{** loaded for the language `#1'. Using the pattern for}%
\typeout{** the default language instead.}%
\else
\language=\csname l@#1\endcsname
\fi
#2}}
\providecommand{\BIBdecl}{\relax}
\BIBdecl

\bibitem{Lee:2010:CF:1837274.1837462}
E.~A. Lee, ``{CPS} {F}oundations,'' in \emph{Proceedings of the 47th Design
  Automation Conference}, ser. DAC '10.\hskip 1em plus 0.5em minus 0.4em\relax
  New York, NY, USA: ACM, 2010, pp. 737--742.

\bibitem{journals/pieee/BanerjeeVMG12}
\BIBentryALTinterwordspacing
A.~Banerjee, K.~K. Venkatasubramanian, T.~Mukherjee, and S.~K.~S. Gupta,
  ``Ensuring safety, security, and sustainability of mission-critical
  cyber-physical systems.'' \emph{Proceedings of the IEEE}, vol. 100, no.~1,
  pp. 283--299, 2012. [Online]. Available:
  \url{http://dblp.uni-trier.de/db/journals/pieee/pieee100.html}
\BIBentrySTDinterwordspacing

\bibitem{Rowe:2008:RSS:1475690.1475895}
\BIBentryALTinterwordspacing
A.~Rowe, K.~Lakshmanan, H.~Zhu, and R.~Rajkumar, ``Rate-harmonized scheduling
  for saving energy,'' in \emph{Proceedings of the 2008 Real-Time Systems
  Symposium}, ser. RTSS '08.\hskip 1em plus 0.5em minus 0.4em\relax Washington,
  DC, USA: IEEE Computer Society, 2008, pp. 113--122. [Online]. Available:
  \url{http://dx.doi.org/10.1109/RTSS.2008.50}
\BIBentrySTDinterwordspacing

\bibitem{DBLP:conf/rtcsa/ChenK07}
J.~Chen and C.~Kuo, ``Energy-efficient scheduling for real-time systems on
  dynamic voltage scaling {(DVS)} platforms,'' in \emph{13th {IEEE}
  International Conference on Embedded and Real-Time Computing Systems and
  Applications {(RTCSA} 2007), 21-24 August 2007, Daegu, Korea}, 2007, pp.
  28--38.

\bibitem{DBLP:conf/rtcsa/KandhaluKLR11}
\BIBentryALTinterwordspacing
J.~Kim, H.~Kim, K.~Lakshmanan, and R.~R. Rajkumar, ``Parallel scheduling for
  cyber-physical systems: Analysis and case study on a self-driving car,'' in
  \emph{Proceedings of the ACM/IEEE 4th International Conference on
  Cyber-Physical Systems}, ser. ICCPS '13.\hskip 1em plus 0.5em minus
  0.4em\relax New York, NY, USA: ACM, 2013, pp. 31--40. [Online]. Available:
  \url{http://doi.acm.org/10.1145/2502524.2502530}
\BIBentrySTDinterwordspacing

\bibitem{Huang:2014:EED:2656045.2656057}
\BIBentryALTinterwordspacing
P.~Huang, P.~Kumar, G.~Giannopoulou, and L.~Thiele, ``Energy efficient dvfs
  scheduling for mixed-criticality systems,'' in \emph{Proceedings of the 14th
  International Conference on Embedded Software}, ser. EMSOFT '14.\hskip 1em
  plus 0.5em minus 0.4em\relax New York, NY, USA: ACM, 2014, pp. 11:1--11:10.
  [Online]. Available: \url{http://doi.acm.org/10.1145/2656045.2656057}
\BIBentrySTDinterwordspacing

\bibitem{conf/ecrts/BaruahBDLMSS12}
S.~K. Baruah, V.~Bonifaci, G.~D'Angelo, H.~Li, A.~Marchetti-Spaccamela,
  S.~van~der Ster, and L.~Stougie, ``The preemptive uniprocessor scheduling of
  mixed-criticality implicit-deadline sporadic task systems.'' in \emph{ECRTS},
  R.~Davis, Ed.\hskip 1em plus 0.5em minus 0.4em\relax IEEE Computer Society,
  2012, pp. 145--154.

\bibitem{Wang2010}
L.~Wang, G.~von Laszewski, J.~Dayal, and F.~Wang, ``Towards energy aware
  scheduling for precedence constrained parallel tasks in a cluster with
  {DVFS},'' in \emph{Cluster, Cloud and Grid Computing (CCGrid), 2010 10th
  IEEE/ACM International Conference on}, May 2010, pp. 368--377.

\bibitem{Beloglazov:2012:ERA:2148243.2148369}
\BIBentryALTinterwordspacing
A.~Beloglazov, J.~Abawajy, and R.~Buyya, ``Energy-aware resource allocation
  heuristics for efficient management of data centers for cloud computing,''
  \emph{Future Gener. Comput. Syst.}, vol.~28, no.~5, pp. 755--768, May 2012.
  [Online]. Available: \url{http://dx.doi.org/10.1016/j.future.2011.04.017}
\BIBentrySTDinterwordspacing

\bibitem{DBLP:journals/fgcs/WuCC14}
C.~Wu, R.~Chang, and H.~Chan, ``A green energy-efficient scheduling algorithm
  using the {DVFS} technique for cloud datacenters,'' \emph{Future Generation
  Comp. Syst.}, vol.~37, pp. 141--147, 2014.

\bibitem{DBLP:journals/fgcs/ParkKF14}
S.~Park, J.~Kim, and G.~Fox, ``Effective real-time scheduling algorithm for
  cyber physical systems society,'' \emph{Future Generation Comp. Syst.},
  vol.~32, pp. 253--259, 2014.

\bibitem{Mobius2014}
C.~Mobius, W.~Dargie, and A.~Schill, ``Power consumption estimation models for
  processors, virtual machines, and servers,'' \emph{Parallel and Distributed
  Systems, IEEE Transactions on}, vol.~25, no.~6, pp. 1600--1614, June 2014.

\bibitem{Rafique:2011:PMH}
M.~M. Rafique and et~al., ``Power management for heterogeneous clusters: An
  experimental study,'' in \emph{IGCC}, Washington, DC, USA, 2011, pp. 1--8.

\bibitem{Davis2011}
\BIBentryALTinterwordspacing
R.~I. Davis and A.~Burns, ``A survey of hard real-time scheduling for
  multiprocessor systems,'' \emph{ACM Comput. Surv.}, vol.~43, no.~4, pp.
  35:1--35:44, Oct. 2011. [Online]. Available:
  \url{http://doi.acm.org/10.1145/1978802.1978814}
\BIBentrySTDinterwordspacing

\bibitem{Masrur2013d}
R.~Schneider, D.~Goswami, A.~Masrur, M.~Becker, and S.~Chakraborty,
  ``\BIBforeignlanguage{english}{{M}ulti-layered scheduling of
  mixed-criticality cyber-physical systems},''
  \emph{\BIBforeignlanguage{english}{Journal of Systems Architecture (JSA)}},
  vol.~59, no. 10-D, 2013.

\bibitem{Abeni2012}
\BIBentryALTinterwordspacing
L.~Abeni, N.~Manica, and L.~Palopoli, ``Efficient and robust probabilistic
  guarantees for real-time tasks,'' \emph{J. Syst. Softw.}, vol.~85, no.~5, pp.
  1147--1156, May 2012. [Online]. Available:
  \url{http://dx.doi.org/10.1016/j.jss.2011.12.042}
\BIBentrySTDinterwordspacing

\bibitem{Chantem2009}
T.~Chantem, X.~Hu, and M.~Lemmon, ``Generalized elastic scheduling for
  real-time tasks,'' \emph{Computers, IEEE Transactions on}, vol.~58, no.~4,
  pp. 480--495, 2009.

\bibitem{narendra2010leakage}
S.~Narendra and A.~Chandrakasan, \emph{Leakage in Nanometer CMOS Technologies},
  ser. Integrated Circuits and Systems.\hskip 1em plus 0.5em minus 0.4em\relax
  Springer, 2010.

\bibitem{rabaey2009low}
J.~Rabaey, \emph{Low Power Design Essentials}, ser. Engineering
  (Springer-11647).\hskip 1em plus 0.5em minus 0.4em\relax Springer, 2009.

\bibitem{Deb2002}
K.~Deb, A.~Pratap, S.~Agarwal, and T.~Meyarivan, ``A {F}ast and {E}litist
  {M}ultiobjective {G}enetic {A}lgorithm: {NSGA}-{II},'' \emph{IEEE
  Transactions on Evolutionary Computation}, vol.~6, no.~2, pp. 182--197, 2002.

\bibitem{Sayyad2013}
A.~Sayyad and H.~Ammar, ``Pareto-optimal search-based software engineering
  (posbse): A literature survey,'' in \emph{Realizing Artificial Intelligence
  Synergies in Software Engineering (RAISE), 2013 2nd International Workshop
  on}, May 2013, pp. 21--27.

\end{thebibliography}

\end{document}